# TOWARDS A GOOD ABS DESIGN FOR MORE RELIABLE VEHICLES ON THE ROADS

Afifa Ghenai[1], Mohamed Youcef Badaoui[1] and Mohamed Benmohammed[1]

[1]LIRE Laboratory, Computer Science Department, University of Constantine, Constantine, 25000, Algeria
afifa.ghenai@gmail.com, joseph-moh@hotmail.com, ben_moh123@yahoo.com

## ABSTRACT

*Nowadays, better driving also means better braking. To this end, vehicle designers must find all failures during the design phase of antilock braking systems which play an important role in automobiles safety. However, mechatronic systems are so complex and failures can be badly identified. So it is necessary to propose a design approach of an antilock braking system which will be able to avoid wheels locking during braking and maintain vehicle stability. This paper describes this approach, in which we model the functional and the dysfunctional behavior of an antilock braking system using stopwatch Petri nets.*

## 1. INTRODUCTION

Mechatronics is an interdisciplinary field which has brought a revolution in the industrial world. According to Industrial Research and Development Advisory Committee of the European Community, mechatronics is the synergetic combination of mechanical engineering and electronic command, with computer systems, used in designing and manufacturing industrial products. Several products in mechanical and electrical engineering areas integrate nowadays a combination of mechanics and electronics [12]. Vehicles are one of the most typical mechatronics products and antilock braking system (ABS) is the first mechatronic product in vehicles which controls the hydraulic pressure of the braking system, so that the wheels do not lock during braking.

However, the simultaneous use of several technologies increases the risk of mechatronic systems dysfunction. That is why reliability becomes one of the major stakes in the last and the coming years. Indeed, mechatronic industries require high level of reliability of their systems especially as failures could cause severe damage and dramatic consequences for the system and the user. So we must include a reliability study during the design phase in order to build systems in which users can put more confidence. Thus, this reliability study must take into account efficiently and in realistic way of time constraints to which the mechatronic systems are subjected which requires the use of a rigorous formalism to model the mechatronic system such as Stopwatch Petri Nets model (SWPN), a powerful tool of design and analysis, particularly adapted to the description of embedded systems [1].

Many timed models are not sufficiently able to model and verify real time applications. Indeed, in these models, time passes in an identical way for each component of the system and the suspension and resumption of task execution cannot be represented [6]. Consequently, models with stop watches in which the concept of clock used in timed models is replaced by a stop watch are proposed. Contrary to a clock, a stop watch preserves its value during the passing of time when it is stopped, then, it is started again. Stopwatch Petri nets (SWPN) are proposed in order to express more temporal behaviors by taking into account the interruption and resumption of tasks, and thus, to give a detailed model of mechatronic systems which allows to identify more dangerous behaviors [1].

The rest of this paper is organized as follows: after giving an overview of related work in section 2, we present our proposed reliability approach and discuss its advantages in section 3. In section 4, we describe a more detailed configuration of a mechatronic system: an antilock braking system using Stopwatch Petri nets. Section 5 describes the application of the method to the antilock braking system and discusses the obtained results. Finally, we conclude the paper and give a brief outlook for future work in section 6.

## 2. RELATED WORK

Several works related to ABS design have been proposed. In [13], Fuzzy Logic Control is suggested to create two different ABS controllers. Authors examine theoretically the braking performance and investigate the influence of vehicle initial speed. In [14], a novel approach to design of ABS controllers is introduced with only input/output measurements of digital sliding mode control and the control algorithm contains the signal of the modeling error. In [15], authors establish the state equation for the dynamics of quarter-car, and present a stable robust sliding mode control based on RBF neural network. Consequently, the reaching phase is eliminated from conventional sliding mode control which guarantees more robustness of the system during the control process. In [16], authors present the rope-less elevator, a technology for high-rise buildings. They analyze the common faults of ABS and propose a rope-less elevator braking system. In order to identify running condition, the proposed method uses hydraulic pressure transducer, disc spring pressure sensor and air gap sensor.

In the following section, we present our proposed approach in which we model the functional and the dysfunctional behavior of an antilock braking system using stopwatch Petri nets.

## 3. THE PROPOSED APPROACH

In order to face the increasing complexity of mechatronic systems and to represent the suspension and resumption of task execution we propose to extract directly feared scenarios which are unknown during the design phase of mechatronic systems from a Stopwatch Petri net model. Feared scenarios approach is proposed by Khalfaoui [2], improved and implemented by Medjoudj [3] and Sadou [4]. Feared scenarios are extracted from a Petri net model without generating the associated reachability graph [5].

### 3.1. Feared scenarios

A scenario can be defined as a beginning, an end and a history which describes the evolution of a system. In reliability study, a feared scenario leads to a catastrophic or dangerous state called feared state. The feared scenario describes how the system leaves a normal state towards this dangerous state. The definition of a scenario is based on the concept of event and relations between the events [4].

**Definition 1. (Event):** We consider a Petri net (P, T, Pre, Post), $M_0$ its initial marking. An event is a particular firing of a transition t ∈ T. the set of events is noted E. From $M_0$, if the transition ti is fired for the $j^{th}$ time, this is the occurrence of the event $e_i^j$

**Definition 2. (Scenario):** A scenario sc, noted sc = (l, $\prec_{sc}$) associated with the Petri net P and the couple $M_0$ and $M_F$ markings, is a set of events l provided with a strict partial order $\prec_{sc}$ defined on the events of l. If for e1, e2 ∈ l : e1 ≺ e2, then the event e1 precedes the event e2 in the scenario sc [4].

### 3.2. Stopwatch Petri nets (Post and Pre initialization)

Stopwatch Petri nets (SWPN) were proposed in order to extend Time Petri nets (TPN) by expressing the behavior of interruptible systems and thus, the suspension and resumption of tasks execution. In a SWPN [7], there are two types of transitions: interruptible and non-interruptible transitions. A mechanism of initialization of the stop watches called post-initialization is used. It is based on the firing of an interruptible transition which puts at zero the stopwatch associated with this transition.

The advantage of stopwatch Petri nets is that they allow a simple graphic formalism where only the initialization of the clocks is modified, the stopwatch is reset, stopped and started [8]. So, interruptible systems can be represented. The principle of SWPN is simpler than IHTPN (Time Petri Nets with Inhibitor Hyper arcs) which use an inhibitor arc to connect a place to an interruptible transition [9]. We can say that SWPN have a combination of two advantages: the Petri nets concision and the analysis power of stopwatch automata.

In the following section, we explain the basic steps of the feared scenarios generation method using a stopwatch Petri net model. The proposed method is presented in [1].

### 3.3. Feared scenarios generation method using stopwatch Petri nets

#### 3.3.1. Principle

Our approach propose a more detailed configuration of the system using Stopwatch Petri nets which allows generating more feared scenarios which cannot be extracted by the preceding feared scenarios approaches. We represent the interruption and resumption of task execution of the ABS components and propose a new version of the algorithm described in [3] and [4].

The stopwatch Petri net model of the ABS is analyzed. We make a back reasoning starting from the feared state and we stop when we reach the first normal states. Then we make a front reasoning from these normal states in order to identify events that lead the system to the dangerous state [1].

#### 3.3.2. Method steps

The proposed method contains four steps which describe how the occurrence of a dangerous event can be identified. Figure1 shows the method principle.

- The first step determines the places whose marking represents a normal functioning state (a nominal state of the system).
- The second step determines target states: a target state can be a feared state (F.S) or states that have direct or indirect causal relations with the feared state (P.F.S).
- The third step consists on making a back reasoning by the use of the inverted stopwatch Petri net of the system. We start from the target state in order to determine the normal functioning states (P.N.S) from which the system can go towards a dangerous behavior. We go back up through the preceding states, until we arrive to normal functioning states called: the conditioners states. The starting points of the next step are these conditioners states.
- The fourth step consists on making a front reasoning starting from the conditioners states in order to determine the possible sequences which lead the system towards a feared behavior. The bifurcations (BIF1) between the normal functioning and the feared state (a bifurcation is a conflict between transitions) give the information of the feared event context. In order to identify the events that lead the system to the feared state, we analyze these conflicts by a marking enrichment.

In order to make a maximal marking enrichment, we introduce the maximum of tokens in the unmarked input places of the potentially fired transitions involved in a conflict. Consequently, the priority transition is fired and the system remains in its normal functioning. In order to make a minimal marking enrichment, we introduce the minimum of tokens in the unmarked input places of the potentially fired transitions. These transitions have a relation with the feared state but are not involved in a conflict [1].

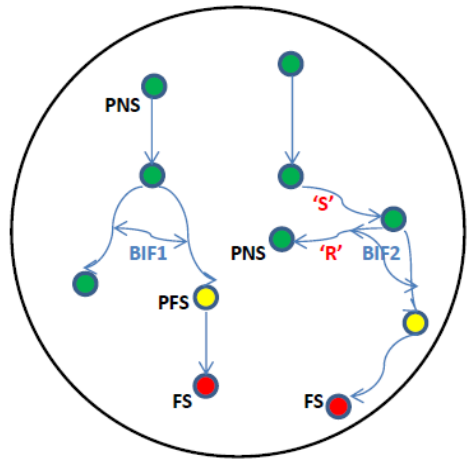

Figure1. Principle of feared scenarios method based on the analysis of stopwatch Petri net model of the system

The use of SWPN model enables us to express temporal behaviors better than TPN model by taking into account the suspension 'S' and resumption 'R' of tasks. This new configuration generates two kinds of bifurcations. The new bifurcations (BIF2) are the conflicts between transitions which represent the non-resumption of interruptible transitions and transitions which represent the resumption 'R' (the normal functioning) of interruptible transitions. The system goes towards a feared state because of the firing of the interruptible transition which cannot be resumpted due to non-respect of time constraints.

Indeed, new feared scenarios which are not found by the preceding feared scenarios approaches can be identified using our proposed method. Consequently, when we make the front reasoning step of the method we must take into account these two kinds of bifurcations. The presence of the new kind of bifurcations is determined by the memorized stopwatch value. In this case, we must modify time constraints in order to make a system reconfiguration [1].

### 3.3.3. Data structures

Input data are changed the proposed feared scenarios generation algorithm. The maximum stop time of a task is added: the input $\alpha_{max}$. The system cannot make a task resumption if the stop time of a task exceeds $\alpha_{max}$. Thus, we add the procedure « check transition ($t_k$) » and we add the condition: (if the transition to be fired is that of the stop, then the stopwatch $\alpha$ starts) in the procedure « fire transition ($t_k$) »

### 3.3.3.1 Input Data

They contain the list of normal tokens ($L_n$) and the list of the initial tokens ($L_i$), $\alpha_{max}$ which enables to define a restart condition of task, and the list of prohibited transitions (Lint).

### 3.3.3.2. Output Data

It is the result of the algorithm: the generated feared scenarios.

### 3.3.3.3. Internal data

- ($L_c$) the list which contains the current tokens.
- The list of prohibited transitions ($L_{int}$).
- ($L_n$) The list which contains transitions of non-initial normal tokens.
- The list of particular transitions ($L_p$) which contains stop transitions ($t_s$) and resumption transitions ($t_r$).
- Stop time of each task ($\alpha$), it enables to calculate the duration of a task suspension.
- The context ($C_i$), $L_c$ is the current list.

Lists of internal data which are generated from $L_c$:

- Fired transitions without conflict with fired transitions (**TfscEc**).
- Potentially fired transitions without conflict (**Tpfsc**).
- Fired transitions in conflict with at least a potentially fired transition (**Tfcpf**).
- Potentially fired transitions in conflict either with fired transitions or with potentially fired transitions (**Tpfc**) [1].

### 3.3.3.4. Procedures

In this paper, we present only some procedures changes because the feared scenarios generation algorithm is so long.

**- Fire a transition ($t_k$):** When the transition is fired, the current list is updated. We remove consumed tokens and add produced tokens. We memorize the events in 'E', and arcs of precedence relation between two events, in 'A' [1].

**If the transition $t_k$ is a transition $t_s$ then**

- We must add $t_s$ in E
- Remove ($t_i$,p) from $L_c$ list and add ($t_i$,$t_s$) in A, for each token ($t_i$,p) necessary to fire $t_s$ ;
- Add a token ($t_s$, ps) in Lc, for each output place $p_s$ of $t_s$.
  $\alpha$++ ;
- If the place $P_k$ is a normal place, add $P_k$ to the list $L_{nni}$

**Else**
- Add $t_k$ in E
- For each token ($t_i$,p) necessary to fire $t_k$ remove ($t_i$,p) from $L_c$ list and add ($t_i$,$t_k$) in A ;
- For each output place $p_s$ of $t_k$, add a token ($t_k$, ps) in Lc.
- If the place $P_k$ is a normal place, add $P_k$ to the list $L_{nni}$

**-Specify a transition ($t_k$):** compare the value of $\alpha$ with $\alpha_{max}$, if $\alpha \leq \alpha_{max}$ then make the resumption of task. However, if $\alpha > \alpha_{max}$ then fire another transition that does not allow the resumption.

**If $\alpha \leq \alpha_{max}$ then**
- Add $t_r$ in E
- Remove $(t_i,p)$ from $L_c$ list and add $(t_i,t_r)$ in A, for each token $(t_i,p)$ necessary to fire $t_r$;
- Add a token $(t_r, p_s)$ in Lc, for each output place $p_s$ of $t_r$.
- If $P_k$ is a normal place, add $P_k$ to $L_{nni}$

**Else ($\alpha > \alpha_{max}$)**
  If $\exists\ t_k$ then
- Remove $t_r$ from the list of sorted transitions.
- Add $t_k$ in E
- Remove $(t_i,p)$ from $L_c$ list and add $(t_i,t_k)$ in A, for each token $(t_i,p)$ necessary to fire $t_k$;
- Add a token $(t_k, p_s)$ in Lc, for each output place $p_s$ of $t_k$.
- If $P_k$ is a normal place, add $P_k$ to $L_{nni}$

**-Sort transition ($t_k$):** We associate respectively the time intervals: $I_k$, .., $I_{k+1}$ to the transitions $t_k$, .., $t_{k+1}$.

$I_{k+1} = [t_{kmin}, t_{kmax}]$. $t_k$ is fired in $T_k$ units of time, with: $t_{kmin}\ T_k\ t_{kmax}$.
$I_{k+1} = [t_{k+1min}, t_{k+1max}]$. The transition $t_{k+1}$ can be fired in $T_{k+1}$ units of time,

with: $t_{k+1min}\ T_{k+1}\ t_{k+1max}$

Strong semantics of time Petri nets is used in our algorithm. It imposes that a transition $t_k$ must be fired at the latest at its date of firing at the latest: $t_{kmax}$.

**Sort transition ($t_k$)**

**For** each transition: $t_k$, $k \in \{1,2..............K....n\}$ /$n \in N$ **do**
**If** $t_{kmin} < t_{k+1min}$ **then** $t_K$ is the first transition to be fired.

**Else**
   If $t_{kmin} > t_{k+1min}$ then $t_{K+1}$ is the first transition to be fired.
   Else
      $t_{kmin} = t_{k+1min}$ then
      If $t_{kmax} < t_{k+1max}$ then $t_K$ is the first transition to be fired.
      Else $t_{kmax} > t_{k+1max}$ then $t_{K+1}$ is the first transition to be fired.

## 4. APPLICATION OF THE APPROACH TO AN ANTILOCK BRAKING SYSTEM

### 4.1. Description

In order to build vehicles in which drivers can put more confidence we include the proposed reliability approach during the design phase of an antilock braking system (ABS). Our approach allows modelling of an ABS using stopwatch Petri nets for a best taking into account of time constraints and thus to guarantee a best level of vehicle reliability. The example chosen presented in [11], is a mechatronic automobile system - Antilock Brake System (ABS) presented in the figure 2.

The brake system is a key component in securing the safety of passengers. Braking performance becomes increasingly important as vehicle speed increases [10]. ABS is a system which prevents the blocking of one or several wheels. Only the nose wheels are controlled using a controller. According to the received information, the computer actuates the valve of the brake system. If the sensors identify that a wheel is locked or that there is a difference between the vehicle speed and the wheel speed, if such a situation occurs, the hydraulic actuators decrease the pressure of the liquid of braking, until the wheel starts to turn or until there is no more difference in measured speed.

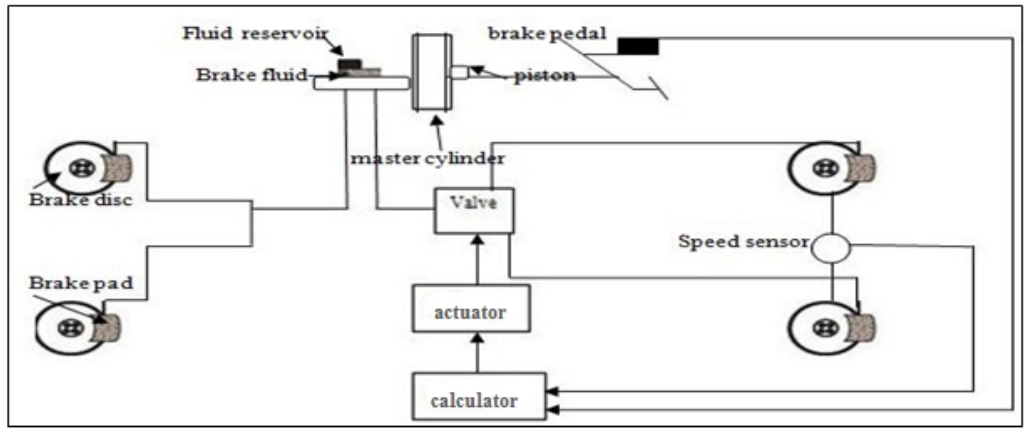

Figure 2. Antilock Braking System (ABS)

## 4.2. System modeling using Stopwatch Petri nets

We propose to model the components of the antilock braking system with Stopwatch Petri Nets (SWPN) in order to represent the suspension and resumption of tasks in each component and to show that there are more feared scenarios obtained by the application of our proposed algorithm.

### 4.2.1. Stopwatch Petri net representation of the common block model

The common block contains the different components of the system: brake pedals, the piston, brake fluid, the liquid reservoir, the calculator (with software), brake pads and brake discs. Figure 3 shows a stopwatch Petri net representation of the common block model.

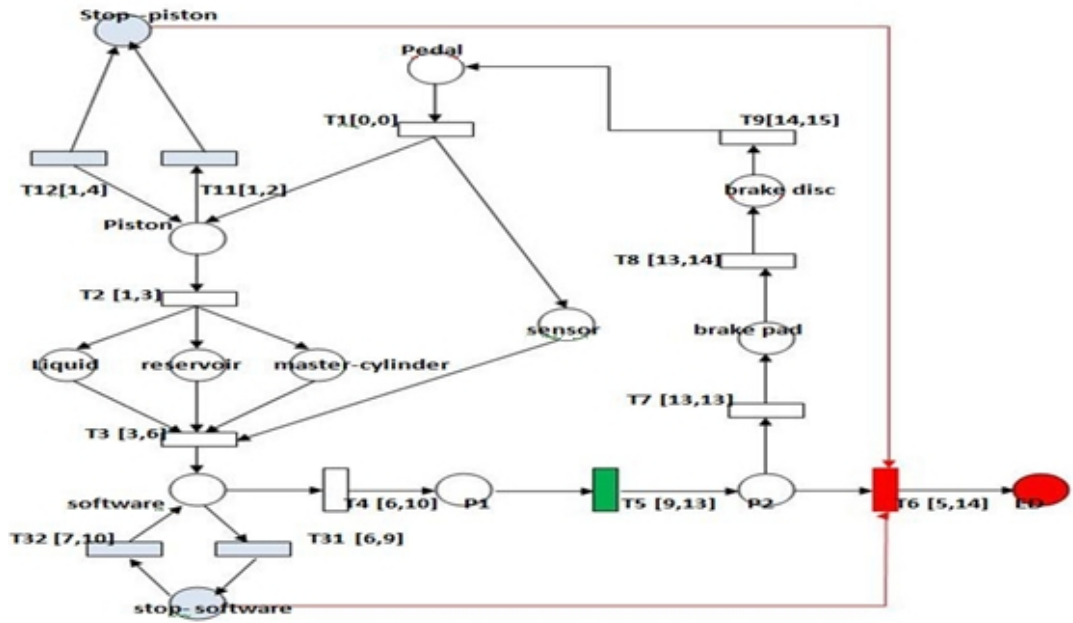

Figure 3. SWPN representation of the common block model

In the proposed approach, we model the functional and the dysfunctional behavior of an antilock braking system. Only, the most important failures are taken into account in the dysfunctional model of ABS.

- Places and transitions in white: corresponds to normal states.
- Places and transitions in blue: corresponds to the suspension and resumption of tasks.
- Places and transitions in red: corresponds to feared states.
- The transition in green: corresponds to the call of ABS object.

In the old ABS model [11], places and transitions in blue are not represented (the suspension and resumption of tasks in each component are not represented). Consequently, the results obtained by the application of the old version of the feared scenarios generation algorithm: there are no feared scenarios. However, our proposed ABS model expresses temporal behaviors better than the old model by taking into account the suspension and resumption of tasks. The advantage of our model is that the more detailed configuration of the antilock braking system enables us to generate feared scenarios because of non-respect of time constraints.

The suspension of the piston task is due to the piston failure. It is represented by the firing of the transition T11 in the time interval [1,2]. The firing of the transition T12 in the time interval [1,4] represents the resumption of task (the reparation of the piston). If the failure duration exceeds this time interval, the system leaves its normal functioning.

The suspension of the software task is due to the software failure. It is represented by the firing of the transition T31 in the time interval [6,9]. The firing of the transition T32 in the time interval [7,10] represents the resumption of task (the reparation of the software). If the failure duration exceeds this time interval, the system leaves its normal functioning.

**4.2.2. Stopwatch Petri net representation of the optional block model**

The optional block contains the component actuator and the valve. If the valve is open, the liquid which goes to brake pedals increases and can cause the wheel locking. So, the actuator, ordered by the computer, prevents the pressure from increasing in the circuit. Figure 4 shows a stopwatch Petri net representation of the optional block model.

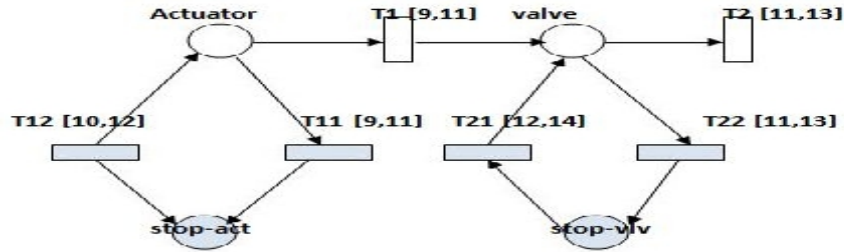

Figure 4. SWPN representation of the optional block model

The suspension of the actuator task is due to the actuator failure. It is represented by the firing of the transition T11 in the time interval [9,11]. The firing of the transition T12 in the time interval [10,12] represents the resumption of task (the reparation of the actuator). If the failure duration exceeds this time interval, the system leaves its normal functioning.

The suspension of the valve task is due to the valve failure. It is represented by the firing of the transition T22 in the time interval [11,13]. The firing of the transition T21 in the time interval [12,14] represents the resumption of task (the reparation of the valve). If the failure duration exceeds this time interval, the system leaves its normal functioning.

At the time of the call of ABS object (the optional block) by the common object (the common block), the actuator sends a request to the valve which will close the brake system. If a component undergoes a failure, the circuit remains open and the system leaves its normal functioning and goes to the feared state: wheel locking.

## 5. APPLICATION OF THE METHOD

In this section, we apply our feared scenarios generation algorithm to the antilock braking system. The application of the four steps of the method using stopwatch Petri nets show more interactions between the different components of this system which allows generating new feared scenarios. These scenarios cannot be generated when we apply the preceding feared scenarios approaches to the old ABS model presented in [11].

Step 1: there are many nominal states of the system: all places in white.
Step 2: we identify the feared state: the wheel locking. It is a target state.
Step 3: using the inverted stopwatch Petri net model of the system, presented in Figure 5, the back reasoning starts from the target state: the wheel locking. We make a back reasoning through all the preceding states, we stop when we reach normal functioning states (conditioner states): stop-piston and stop-software.

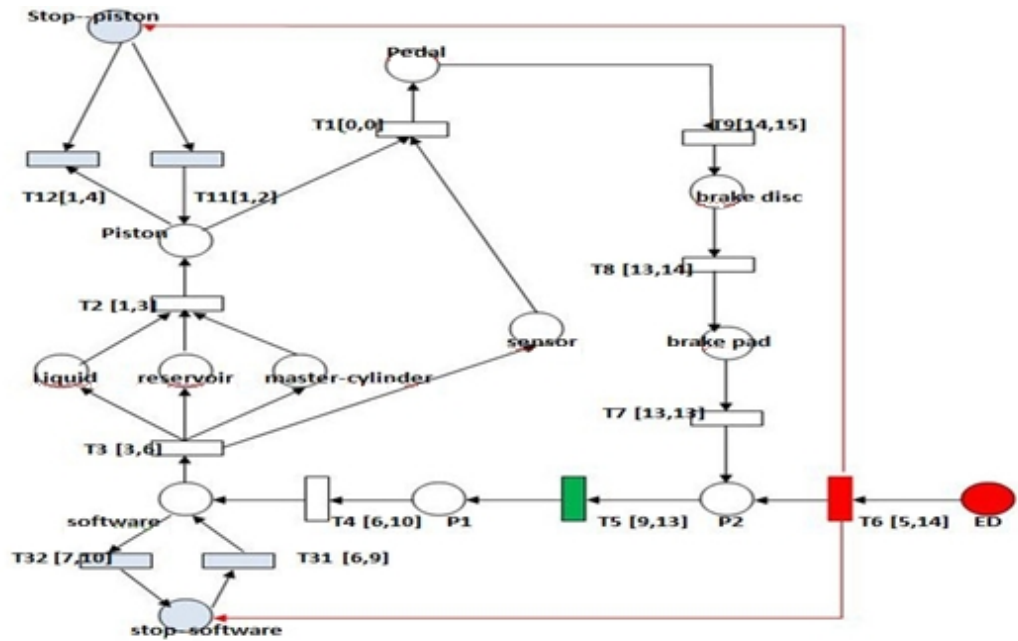

Figure 5. Inverted Petri net representation of the common block

Step 4: the front reasoning begins from the conditioner states: stop-piston, stop-software (or: stop-vlv, stop-act). These places which are input places of the transition T6, are marked and this is the cause of the second kind of bifurcation: BIF2.

If the place stop-piston is marked (this marking causes a conflict between the transitions T12 and T6). A task is then interrupted: the piston task. This suspension is memorized during the front reasoning. If the duration of this suspension exceeds its time interval, we must memorize the stopwatch value. The system leaves then its normal functioning because the transition T12 (the resumption of task) is not fired. The place ED (a feared state: the wheel locking) is then marked by the firing of the transition T6. Thus, time constraints must be modified in order to avoid the drift towards the wheel locking.

Since our algorithm is too long, we present in this paper, only some steps of back reasoning and front reasoning.

**Initial step**

Lc=Li={(i,ED)},Lint={ },Le={ },E={i},A={ },inc=1,C={(Lc,Lint,E,A ,Le)}.

**Step 1:** C is not empty
C={(Lc , Lint,E,A ,Le)}={( i, ED),{ }, {i},{ },{ }}
C becomes empty. Go to step 2;

**Step 2:**
The only fired transition is t6.
TfscEc = {t6} ; Go to step 3 ;

**Step 3**:
Lc does not contain only tokens belonging to Ln (ED does not belong to Ln) from where: Go to step 4 ;

138

**Step 4:**
TfscEc (= {t6}) is not empty from where: tk = t6
Fire transition (t6) : E = {i, t6}
Lc = { } and A = {(i, t6)} ; Lc = {(t6, stop-piston)} ; Go to step 2**.**

**Front reasoning**

We have a token in the place piston. Li = {(i, piston)} = Lc.

**Initial step :**
$_{max}$ = $t_{22\ max}$ =13  (ABS object)

**Step 1:**
C is not empty ,C={(Lc , Lint,E,A ,Le)}={( i, piston),{}, {i},{},{}} ; C becomes empty.
Go to step 2 ;

**Step 2:**
TfscEc = {t11, t2} ; Go to step 4.

**Step 4:**
To sort transitions ($t_k$) ;
tk= t11 the first transition de TfscEc ;
To memorize context (t11)
Lint = { }
Add t11 to Lint : Lint = {t11}
C = (Lc = {(i, piston)}, Lint = {t11}, E = {i}, A = {}, Le = {}).
To erase the contents of Lint: Lint = { }
Fire Transition (t11): E = {i, t11}
Lc = {(t11, stop-piston)} ;
  ++
Go to step 2

**Step 2:**
Tfcpf={t12 } ;tpfc={t6}
Go to step 5.

**Step 5**:
To enrich Marking1 (t6): Initially L = { }
The only transition in conflict with t12 is t6. We add a token (e1, P2).
Thus, $L_e$ = {(e1, P2)}.
This marking enrichment is coherent because: M(P2) + M(ED) = 1.
Lc = {(i, stop-piston), (e1, P2)}, Le = {(e1, P2)}
To memorize Context (t12)
    0
To specify transition ($t_k$)
If     13 then
E = {i,t11,t12$_r$}
Lc = {(e1, P2) }. A = {(i, t12)}
Lc = {(t12,piston) , (e1, P2)  }
Add stop-piston à $L_{nni}$.
Else   >13

139

E = {i, t11,t6}
Lc = {(e1,P2) ,(I, stop-piston)}.
Lc = {(t6,ED)} ,Lint = { }, E = {i, t11,t6}. Go to step 2 ;

**Step 2:**
All the lists are empty.

**Step 3:**
The stop criterion is satisfied. No transition is fired.
Go to step 8.

**Step 8:**
The built partial order is defined by:
E1 = {i, t11, t12}, A1 = {(i, t11), (t11, t12), (t12, f1)},
E2= {i, t11, t6}, A2 = {(i, t11), (t11, t6), (t6, f2)}, Le = {(e1, P2)}.
Go to step 1.

**Step 1:**
C = (Lc = {(i, piston)}, Lint = {t11}, E = {i}, A = {}, Le = {}. C becomes empty…

Our approach propose a more detailed configuration of the system using Stopwatch Petri nets which allows generating more feared scenarios which cannot be extracted by the preceding feared scenarios approaches and when we use the old ABS model presented in [11]. There are more interactions between the ABS components due to the best expression of temporal behaviors, especially, the representation of the suspension and resumption of tasks.

The application of our algorithm to ABS system enabled us to generate four scenarios that lead to the feared state (the wheel locking). We present also the interactions between the two objects which lead to this feared state. We represent the generated scenarios by partial orders. Each partial order is a directed graph (E, A), the nodes E are transition firings and the arcs A are pairs $(t_i, t_j)$, $t_i$ and $t_j$ are transition firings and $t_i$ precedes $t_j$.

**1$^{st}$ feared scenario:**

E2= {i, t11, t6}, A2 = {(i, t11), (t11, t6), (t6, f2)}, Le = {(e1, P2)}.

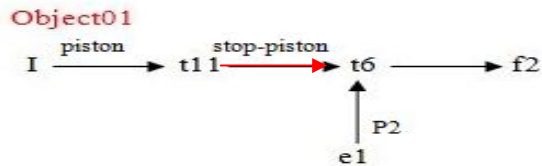

**2$^{nd}$ feared scenario:**

E3= {i, t2, t3, t31, t6}, A2 = {(i, t2), (t2, t3), (t3, t31), (t31, t6) (t6, f4)}, Le = {(e3, P2)}.

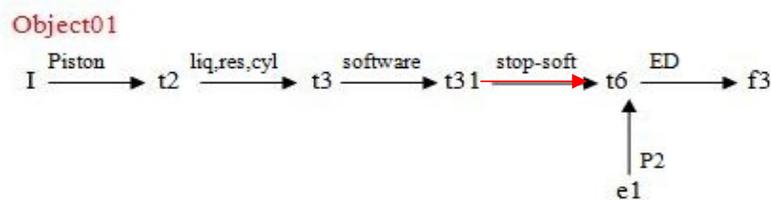

### 3$^{rd}$ feared scenario:

E5 = {i, t2, t3, t4, t5, t6}, A1 = {(i, t2), (t2, t3), (t3, t31), (t31, t32) (t32, f5)}, Le = {(e4, stop-piston), (e5, stop-software)}. (Stop in the actuator)

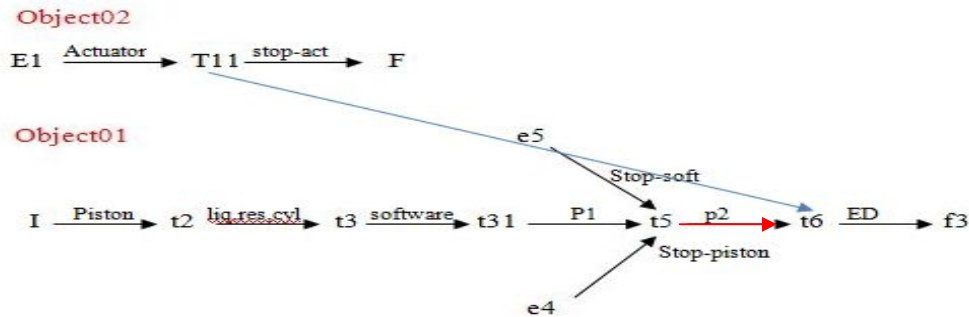

### 4$^{th}$ feared scenario:

E5 = {i, t2, t3, t4, t5, t6}, A1 = {(i, t2), (t2, t3), (t3, t31), (t31, t32) (t32, f5)}, Le = {(e4, stop-piston), (e5, stop-software)}. (Stop in the valve)

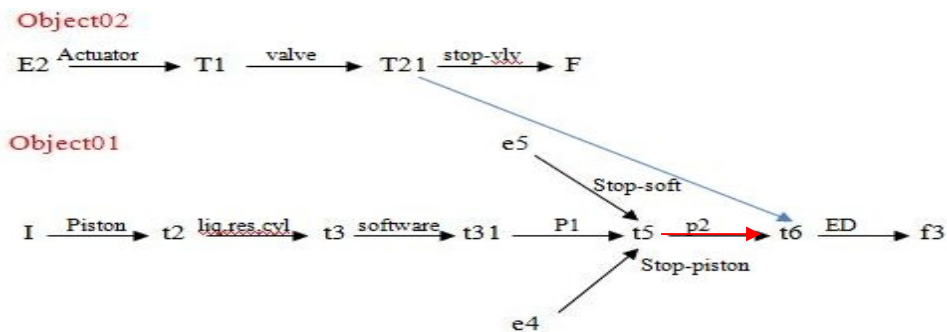

## 6. CONCLUSIONS

In this paper, we proposed a reliability approach during the design phase of an antilock braking system (ABS) in order to avoid failures which could cause dramatic consequences for the system and the user. Our approach allows modelling of an ABS using stopwatch Petri nets for a best taking into account of time constraints and thus to guarantee a best level of vehicle reliability. This approach proposes a more detailed configuration of the ABS and allows generating more feared scenarios which cannot be extracted by the preceding feared scenarios approaches and when we use the old ABS model. There are more interactions between the ABS components due to the best expression of temporal behaviors using stopwatch Petri nets, especially, the representation of the suspension and resumption of tasks.

Consequently, we can find a new kind of conflicts between transitions and thus, new feared scenarios (dangerous behaviors) which are generated because of non-respect of time constraints.
In the future, we plan to propose an extended hybrid analysis of the antilock braking system by taking into account of continuous dynamics of its components. The continuous dynamic induces a firing order of transitions. This order depends on the nature of the dynamic and allows generating more feared scenarios.

**Authors**

**Afifa Ghenai** is with the department of computer science, University of Constantine, Algeria. She is a member of LIRE laboratory. Her research domains are formal methods and mechatronic systems.

**Mohamed Youcef Badaoui** has a master degree in computer science from the University of Constantine, Algeria in 2011. His research domain is real time systems.

**Mohamed Benmohammed** is a professor at the University of Constantine, Algeria. He is the head of the AS group of LIRE laboratory. His research field is embedded systems.